# Suppléance perceptive par électro-stimulation linguale embarquée : perspectives pour la prévention des escarres chez le blessé médullaire paraplégique

Olivier Chenu
Laboratoire TIMC-IMAG, UMR UJF
CNRS 5525
Faculté de Médecine de Grenoble
38706 La Tronche cédex
+33 4 56 52 00 05
Olivier.Chenu@imag.fr

Nicolas Vuillerme, Alexandre Moreau-Gaudry, Anthony Fleury, Jacques Demongeot, Yohan Payan
Laboratoire TIMC-IMAG, UMR UJF CNRS 5525
Faculté de Médecine de Grenoble
38706 La Tronche cédex

Prénom.Nom@imag.fr

**RESUME**

Ce travail présente une étude de faisabilité de l'utilisation d'un dispositif biomédical de suppléance perceptive par électro-stimulation linguale embarquée pour la prévention des escarres chez le blessé médullaire paraplégique. Exploitant le paradigme de « suppléance perceptive », le principe de ce dispositif consiste à (1) identifier des zones de souffrance tissulaire à partir de l'analyse automatique des pressions enregistrées au niveau des fessiers et (2) fournir une stimulation au niveau de la langue d'une mobilisation posturale à adopter pour soulager cette surpression et ainsi éviter la formation d'une escarre. La performance de ce dispositif a été évaluée chez 6 sujets sains. Nos résultats montrent que les sujets parviennent à intégrer une information artificielle linguale pour mobiliser leur buste selon une information directionnelle donnée afin de soulager les zones de surpression. Ces résultats permettent d'envisager l'utilisation de ce dispositif de suppléance perceptive par électro-stimulation linguale embarquée pour la prévention des escarres chez le blessé médullaire paraplégique. Il s'agit désormais d'en évaluer, quantitativement et qualitativement, l'apport effectif sur des populations de patients paraplégiques et de personnes âgées et/ou déficientes en situation.

**Descripteurs de sujet et catégories**

Sciences de la vie et médecine – *Santé*.

**Termes Généraux**

Expérimentation

**Mots-clés**

Suppléance perceptive - Prévention - Escarre – Paraplégique

## 1. INTRODUCTION

### 1.1 Contexte général

Un des axes de recherche du Laboratoire TIMC-IMAG a trait au développement de technologies innovantes à la fois pour les capteurs intégrés multimodaux, embarqués et communicants et la rééducation, la correction et la suppléance fonctionnelle du handicap. Deux objectifs complémentaires sont poursuivis.

Sur un plan technologique d'une part, il s'agit de développer des dispositifs biomédicaux de suppléance perceptive / motrice / cognitive, ergonomiquement et esthétiquement acceptables.

Sur un plan clinique, d'autre part, il s'agit de mettre au point et de valider différentes applications dédiées au handicap et faire en sorte qu'elles puissent effectivement bénéficier à la personne déficiente en condition de validité écologique.

Notons que ces deux objectifs s'inscrivent dans une volonté de contribuer à l'amélioration de la qualité de vie des personnes atteintes de déficiences, et, plus globalement, de contribuer à la réduction des dépenses de santé [1,2].

Plus particulièrement, a été retenue la problématique de prévention des escarres chez le blessé médullaire paraplégique qui, compte tenu de leur prévalence, des conséquences sur la qualité de vie de personnes atteintes et des coûts de santé, constitue une priorité de santé publique.

### 1.2 Les escarres chez le blessé médullaire paraplégique

*1.2.1 Enjeux médicaux et socio-économiques*

Une escarre est une plaie souvent profonde qui se forme aux zones d'appuis des personnes immobilisées suite à la compression prolongée des tissus et à leur privation en sang [3]. Assises dans leur fauteuil, les personnes lésées médullaires paraplégiques, et dans une moindre mesure les personnes âgées, ne perçoivent plus la pression exercée au niveau de leur région fessière et ne peuvent donc plus agir de manière adaptée afin de soulager les zones de surpression en vue de prévenir la formation d'une escarre [4,5]. Ces personnes développent ainsi des escarres (Figure 1), localisées principalement dans les régions ischiatiques.

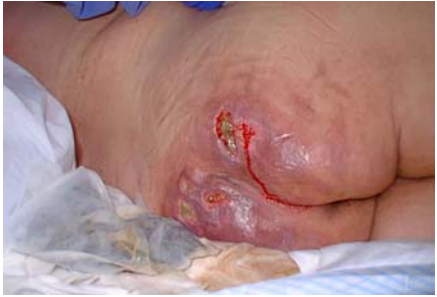

**Figure 1,** Photo d'escarres dans les régions fessières

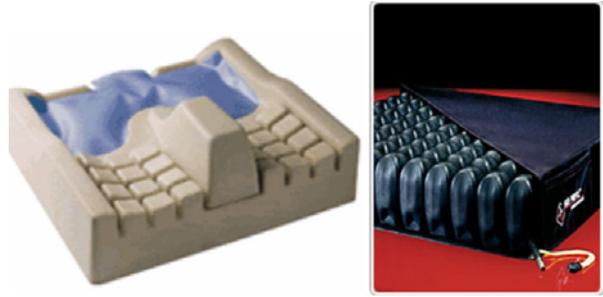

**Figure 2,** Photo de coussins de distribution des pressions

En 1996, parmi les tétraplégiques de 35 services et centres de rééducation fonctionnelle en France, Suisse, et Belgique, 63% avaient présenté une escarre entre l'accident et la fin du premier séjour de rééducation [5]. Les escarres représentaient la 3$^{ème}$ cause de réhospitalisation de ces patients [5]. En 2001, 34 à 46% des blessés médullaires présentaient une escarre dans les 2 ans à distance de l'accident [3]. Outre ses graves retentissements fonctionnels, psychologiques et sociaux, l'escarre peut entraîner ou participer à l'aggravation de pathologies (septicémies) mettant en jeu le pronostic vital [6].

Par ailleurs, même si certaines sources avancent des chiffres de plusieurs milliards d'euros concernant les 300 000 personnes souffrant d'escarres en France [7], les coûts, et plus largement les incidences économiques des escarres, sont encore mal connus. Il est cependant clairement établi que (1) le traitement médical et chirurgical de l'escarre est lourd, long et coûteux, (2) la non application de pratiques de prévention est génératrice de surcoûts importants, en termes d'allongement des durées de séjours, de morbidité accrue ou d'alourdissement de la charge en soins [3].

En définitive, les conséquences des escarres, de nature physique, psychologique, économique et sociale, en font un problème majeur de santé publique.

### 1.2.2 Solutions actuelles

Devant de telles conséquences, différentes solutions préventives ont été proposées chez les patients paraplégiques, dont, en particulier, le contrôle de la pression exercée sur les tissus mous à l'interface saillies-osseuses/fauteuil afin de prévenir l'ischémie et donc la formation d'une escarre [8].

Pour contrôler la pression à cette interface, des coussins et matelas « anti-escarres » ont d'une part été développés et sont maintenant communément utilisés dans les services de soins (Figure 2) [9]. Néanmoins, ces supports s'avèrent insuffisants et les techniques de transfert de poids constituent toujours le meilleur moyen de prévention. Soulèvements, changements de position, modifications des appuis sont à réaliser régulièrement afin d'assurer une bonne répartition des pressions et ainsi prévenir l'ischémie [10].

Des dispositifs permettant la détection des régions en surpression trop longuement exposées, au moyen de nappes de capteurs de pression disposées sous les fesses et reliées à un ordinateur, ont d'autre part été développés (Figure 3). Ils renvoient cette information au patient sous la forme d'un signal auditif ou visuel, permettant ainsi l'identification des zones à risques à partir desquelles le patient pourra corriger sa posture. Cependant, de notre point de vue, les dispositifs actuels ne sont pas ergonomiquement acceptables pour le patient (i.e. non dérangeants dans l'accomplissement de ses tâches perceptives ou motrices concurrentes) En particulier, ces dispositifs sont lourds et encombrants, car physiquement reliés à un ordinateur. De plus, ils relayent l'information par le biais du canal auditif ou visuel, ce qui empêche l'utilisateur de percevoir et d'utiliser les signaux extérieurs, visuels ou auditifs pour l'exécution de tâches concurrentes.

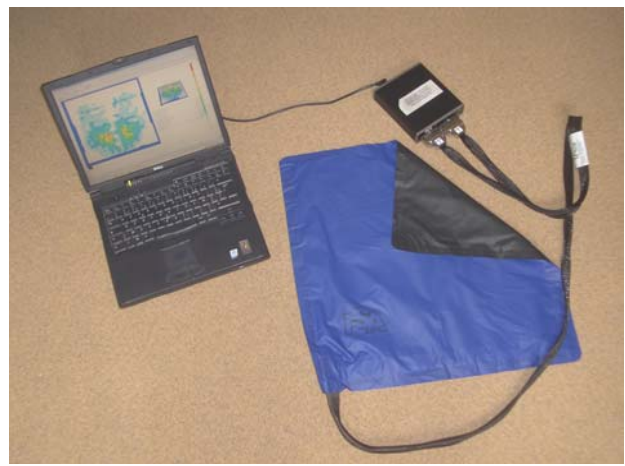

**Figure 3,** Nappes de pressions
(Vista Medical Ltd.)

Dans ce contexte, proposer un nouveau dispositif biomédical intelligent, embarqué sur le paraplégique et sur son fauteuil roulant, visant à renvoyer directement au patient (en condition de validité écologique) les informations de pression mesurées sur l'assise, pourrait avoir très rapidement un impact fort non seulement sur la qualité de vie des personnes atteintes, mais aussi sur l'économie de la santé.

## 2. DISPOSITIF DE PREVENTION DES ESCARRES

### 2.1 Exploitation du paradigme de suppléance perceptive

Le caractère particulièrement innovant de notre recherche repose sur l'exploitation du paradigme de « substitution sensorielle » [11] ou de « suppléance perceptive » [12] qui consiste à fournir de l'information dévolue à une modalité sensorielle par le biais d'une autre modalité [13].

D'une manière générale, un système de suppléance perceptive se compose de trois éléments distincts : des capteurs (1) permettent la conversion d'une forme d'énergie (photonique, sonore, mécanique ou autre) en signaux interprétables par un système de couplage (2) qui est responsable de l'activation coordonnée de stimulateurs (3). Si l'on considère l'indépendance relative de ces trois éléments, ainsi que la grande diversité des types de capteurs et d'actionneurs disponibles à l'heure actuelle, de nombreux systèmes de suppléance perceptive existent déjà [12] ou sont envisageables.

Dans une perspective d'utilisation en recherche appliquée en ingénierie pour la santé, telle que nous l'envisageons, le choix ou le développement d'un dispositif de suppléance perceptive doit non seulement être guidé par les contraintes de son utilisation, mais doit aussi prendre en compte les besoins spécifiques, ainsi que les capacités sensori-motrices, cognitives et fonctionnelles de son utilisateur. Dans ce cadre, nous avons centré nos travaux sur un dispositif de suppléance perceptive tactile constitué d'une matrice d'électrodes mise en contact avec la surface supérieure de la langue (le Tongue Display Unit, « TDU »). En effet, lorsque la plupart des centres nerveux et moteurs rachidiens ont été touchés par un handicap, la langue se trouve très souvent être une des rares structures encore mobile et innervée, puisque les nerfs crâniens qui l'innervent (nerfs grand hypoglosse, trijumeau, glosso-pharyngien et facial) sont généralement épargnés. La version originale développée par Paul Bach-y-Rita [14] pour des applications de substitution visuelle était constituée d'une matrice de 12×12 électrodes reliées par voie filaire à une électronique déportée (Figure 4).

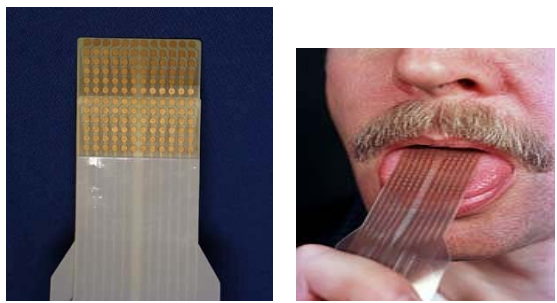

**Figure 4,** Photo du TDU filaire (matrice de 12 × 12 électrodes) [14]

Pour rendre ce dispositif ergonomiquement et esthétiquement acceptable, nous avons mis au point une version télémétrique constituée d'une matrice 6×6 électrodes, de taille plus réduite, collée sur la surface inférieure d'une prothèse palatine orthodontique incluant un circuit électronique, une antenne radio et une pile (Figure 5).

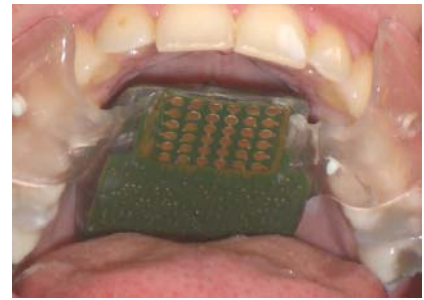

**Figure 5,** Photo du TDU télémétrique (matrice de 6 × 6 électrodes)

### 2.2 Mise en œuvre

Nous avons mis au point un dispositif de suppléance perceptive par électro-stimulation linguale pour la prévention de la formation d'escarres (Payan et al., 2006) [15] fondé sur :

(1) l'identification des zones de souffrance tissulaire à partir de l'analyse automatique des pressions enregistrées au niveau des fessiers et

(2) l'électro-stimulation linguale d'une direction de mobilisation posturale à adopter pour corriger cette surpression (Figure 6).

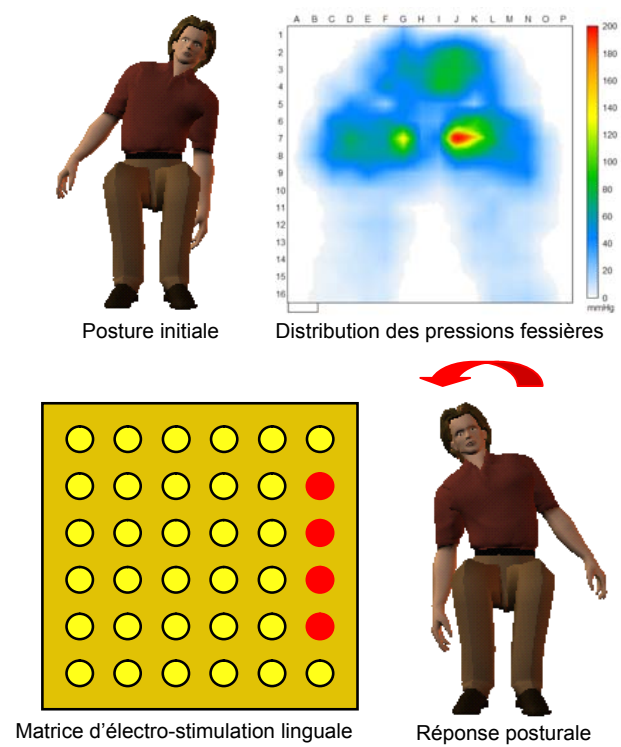

Posture initiale · Distribution des pressions fessières

Matrice d'électro-stimulation linguale · Réponse posturale

**Figure 6,** Principe du dispositif de prévention des escarres

Moreau-Gaudry et al avaient procédés à une étude similaire avec la version filaire du TDU et une nappe de pression développée pour l'occasion.. Les ordres directionnels ètaient aléatoires et, par conséquent, indépendants des pressions précédement enregistrées et le codage employé ne semblait pas adapté à un signal de danger.

## 2.3 Étude de faisabilité

L'objectif de cette étude était d'évaluer si et dans quelle mesure des sujets jeunes sains étaient capables d'intégrer une information artificielle linguale pour mobiliser leur buste selon une information directionnelle donnée afin de soulager les zones de surpression.

Six sujets jeunes sains (âge = 27.2 ± 3.7 ans ; poids = 78.2 ± 9.7 kg; taille = 181.7 ± 9.1 cm) ont volontairement participé à cette étude.

Les sujets, dispositif d'électro-stimulation linguale télémétrique en bouche (Figure 5), étaient confortablement assis sur une chaise. Sur son assise était posée une fine nappe souple munie de capteurs de pression résistifs piézo-électriques (FSA Seat 32/63, Vista Medical Ltd.) permettant le recueil des distributions de pressions fessières.

La procédure expérimentale a successivement consisté à :

(**1**) enregistrer la carte de pression fessière du sujet assis, sur une période variant aléatoirement de 15 à 45 secondes, afin de localiser une zone de surpression (Figure 6, en haut à droite);

(**2**) déterminer la zone, atteignable par un mouvement du buste en avant, en arrière, à droite ou à gauche, susceptible de réduire le plus cette surpression ;

(**3**) activer 4 électrodes des bords antérieur, postérieur, droit ou gauche de la matrice d'électro-stimulation linguale, selon l'algorithme suivant. Si la zone de surpression déterminée initialement était susceptible de disparaître avec un déplacement du buste vers l'avant, l'arrière, la droite ou la gauche, les 4 électrodes situées sur les bords arrière, avant, gauche et droit étaient respectivement activés (Figure 5, en bas à gauche).

Après avoir identifié la région de la matrice activée, le sujet avait pour consigne de mobiliser son buste selon la direction opposée à celle indiquée par l'électro-stimulation linguale (Figure 6, en bas à droite) [16].

(**4**) enregistrer la nouvelle carte de pression fessière du sujet assis afin de la comparer avec celle enregistrée initialement.

Les sujets ont réalisé 2 sessions de 20 essais, pour un total de 40 essais.

Un score unitaire ou nul a été attribué à chaque essai selon que la réponse posturale concordait ou non avec l'information linguale électro-stimulée. Par exemple, après une electrostimulation à droite, la mobilisation est estimée correcte si le centre de pression s'est déplacé vers la gauche (à 45 degrés près)

Le pourcentage de réponses posturales adaptées à la stimulation linguale est de 95.5%. Les échecs ètaient souvent dus à une mauvaise compréhension du signal et, plus rarement, à sa non-détection. Ce résultat suggère que les sujets jeunes sains ont été capables d'intégrer une information artificielle linguale pour mobiliser leur buste selon une information directionnelle donnée afin de soulager les zones de surpression.

## 3. CONCLUSION

La prévention des escarres chez le blessé médullaire paraplégique représente un enjeu de société significatif qui n'a pas aujourd'hui de solutions thérapeutiques fiables et satisfaisantes.

Ce travail présente une étude de faisabilité de l'utilisation d'un dispositif biomédical embarqué visant à répondre cliniquement à cette priorité de santé publique.

Exploitant les paradigmes de « substitution sensorielle » [11] ou « suppléance perceptive » [12], le principe de ce dispositif de prévention des escarres consiste à (1) identifier des zones de souffrance tissulaire à partir de l'analyse automatique des pressions enregistrées au niveau des fessiers et (2) fournir une stimulation au niveau de la langue d'une mobilisation posturale à adopter pour soulager cette surpression et ainsi éviter la formation d'une escarre [15]. La performance de ce dispositif a été évaluée chez 6 sujets sains.

Les résultats montrent que les sujets parviennent à intégrer une information artificielle linguale pour mobiliser leur buste selon une information directionnelle donnée afin de soulager les zones de surpression. Bien sûr, ce dispositif nécessite encore quelques avancées : une évolution algorithmique pour prendre en compte les pressions plus en détail et une évolution technologique afin de s'affranchir de l'ordinateur en intégrant la partie couplage dans un boitier électronique et de miniaturiser encore le TDU. Ces deux derniers points sont tout à fait envisageables aujourd'hui et sont en cours de discussion avec nos partenaires industriels. En attendant, les résultats présentés dans ce papier permettent d'envisager l'utilisation de ce dispositif de suppléance perceptive par électro-stimulation linguale embarquée pour la prévention des escarres chez le blessé médullaire paraplégique. Or cette population a, par définition, des fonctions sensori-motrices altérées par rapport aux sujets sains. Il s'agit donc désormais d'évaluer, quantitativement et qualitativement, l'apport effectif de ces deux dispositifs sur des populations de patients paraplégiques en situation.

## 4. REMERCIEMENTS

## 5. RÉFÉRENCES